\documentclass[aps,pra,twocolumn,showpacs]{revtex4}

\usepackage[T1]{fontenc}
\usepackage{amsmath,amsfonts,amssymb,amsthm,bbm}
\usepackage{graphicx}
\usepackage{color}

\allowdisplaybreaks

\begin{document}

\title{Quantum metrology with non-equilibrium steady states of quantum spin chains}

\author{Ugo Marzolino and Toma\v z Prosen}

\affiliation{Department of Physics, Faculty of Mathematics and Physics, University of Ljubljana, Jadranska 19, SI-1000 Ljubljana, Slovenia}

\date{\today}

\begin{abstract}
We consider parameter estimations with probes being the boundary driven/dissipated non-equilibrium steady states of $XXZ$ spin $1/2$ chains. The parameters to be estimated are the dissipation coupling and the anisotropy of the spin-spin interaction. In the weak coupling regime we compute the scaling of the Fisher information, i.e. the inverse best sensitivity among all estimators, with the number of spins. We find superlinear scalings and transitions between the distinct, isotropic  and anisotropic, phases. We also look at the best relative error which decreases with the number of particles faster than the shot-noise only for the estimation of anisotropy.
\end{abstract}

\maketitle

\section{Introduction}

Metrology, i.e. the ability to perform precise measurements, is central for technological and experimental developments. Among the most advanced metrological schemes, there are thermometry \cite{Spietz2003,Weld2009,Brunelli2011,Zhou2011,Muller2010,Sanner2010,Stace2010,Leanhardt2013,Marzolino2013,Weng2014}, magnetometry \cite{Savukov2005,Loretz2013,Fang2013,Aiello2013,Eto2013,Steinke2013,Waxman2014,Puentes}, and interferometry \cite{Luis2004,Higgins2007,Boixo2007,Boixo2008,Roy2008,Higgins2009,Pezze2009,
Tilma2010,Gross2010,Riedel2010,Benatti2010,Benatti2011,Argentieri2011,Napolitano2011,Rivas2012,Wolfgramm2013,Berrada2013,Benatti2013,Benatti2014}.
Recently, also estimations of hamiltonian parameters have been suggested \cite{DiFranco2009,Burgarth2009,Wiesniak2010,Burgarth2011}.
In the quantum domain, either thermal equilibrium or pure states have been mostly employed. Moreover, real experiments are always affected by unavoidable noise, due to the interaction with the environment, that can be described by markovian master equations under controllable approximaitons \cite{BreuerPetruccione2002,BenattiFloreanini2005}. Dissipation have been shown to be greatly detrimental for interferometry \cite{Kolodynski2010,Escher2011}. On the other hand, estimations of parameters of noisy dynamics have been proposed in \citep{Bellomo2009,Bellomo2010a,Bellomo2010b,Monras2011,Braun2011,Alipour2014}.

In the present paper, we consider a spin $1/2$ chain with $XXZ$ spin interaction driven by local noise at its ends. The unique asymptotic state of the corresponding master equation has been recently derived in terms of matrix product operators \cite{Prosen2011a,Prosen2011b,Prosen2012,Karevski2013,Prosen2013a,Prosen2013b,Prosen2014}, resulting in a non-equilibrium steady state (NESS). Such NESS depends on the parameters describing the dynamics and the properties of the environments at the ends of the chain. Based on this dependence, we propose to use the NESS as a probe to estimate the above parameters.

We quantify the performance of the parameter estimations with the Fisher information \cite{Helstrom1976,Holevo1982,Braunstein1996,Paris2009}, which gives principally the best achievable sensitivity. It is typically important to study the scaling of the Fisher information with the number of resources: faster the Fisher information grows, either smaller devices are required for constant sensitivities or more precise estimations can be performed at fixed sizes. Most frequently, classical devices show a linear scaling of the Fisher information, and thus a linear decrease of the best absolute sensitivity, known as shot-noise. We find superlinear scalings of the Fisher information with respect to different parameters and in different regimes.
In particular, we find a phase transition between power-law scaling of Fisher information in the regime of easy-plane interactions to super-exponential scaling in the regime of easy-axis interactions, for the perturbative range of environment coupling. Within this perturbative analysis, we need to consider the relative error wich decreases faster than linearly in the particle number only for the estimation of the anisotropy parameter for isotropic and easy-plane interactions. In the latter case, the rate of growth of Fisher information is a no-where continuous function of the anisotropy parameter.

\section{The system}

We focus on the following markovian dissipative dynamics of one-dimensional $n$-spin chains:
\begin{eqnarray}
\frac{d}{dt}\rho(t) & = & -i\left[\frac{\Omega}{2}M_z+J H_\textnormal{XXZ},\rho(t)\right] \nonumber \\
& & +\lambda\sum_{k=1}^4\left(L_k\rho(t)L_k^\dag-\frac{1}{2}\left\{L_k^\dag L_k,\rho(t)\right\}\right),
\label{master.eq}
\end{eqnarray}
where $M_z=\sum_{j=1}^n\sigma_j^z$ is the total magnetization along the $z$ direction,

\begin{equation}
H_\textnormal{XXZ}=\sum_{j=1}^{n-1}(\sigma_j^x\sigma_{j+1}^x+\sigma_j^y\sigma_{j+1}^y+\Delta\sigma_j^z\sigma_{j+1}^z)
\end{equation}
is the hamiltonian of the XXZ spin chain,

\begin{equation}
L_{1,2}=\sqrt{\frac{1\pm\mu}{2}}\sigma_1^{\pm}, \qquad L_{3,4}=\sqrt{\frac{1\mp\mu}{2}}\sigma_N^{\pm}
\end{equation}
are the Lindblad noise driving channels, and $\sigma_j^\alpha$ are the Pauli matrices of the $j$-th spin. Master equations with local Lindblad operators, i.e. each environment interacting with a single particle as in \eqref{master.eq}, can be derived from microscopic models of system-environments interaction following \cite{Wichterich2007} provided $\Omega\gg J$, with a concrete example studied in \cite{Oxtoby2009}. The local hamiltonian generator $-i[M_z,\cdot]$ commutes with the other terms in \eqref{master.eq}, namely $-i[H_\textnormal{XXZ},\cdot]$ and $L_k\cdot L_k^\dag-\frac{1}{2}\left\{L_k^\dag L_k,\cdot\right\}$. Therefore, the NESS $\rho_\infty\equiv\lim_{t\to\infty}\rho(t)$ does not depend on the presence of the generator $-i[M_z,\cdot]$, and formally equals the NESS derived in the absence of such generator \cite{Prosen2011a,Prosen2011b,Prosen2013b}, which is unique:

\begin{eqnarray} \label{NESS.pert}
&& \rho_\infty=2^{-n}\bigg(\mathbbm{1}+ i\frac{\lambda}{2J}\mu(Z-Z^\dag) \nonumber \\
&& +\frac{\lambda^2}{8J^2}\left(\mu[Z,Z^\dag]-\mu^2\left(Z-Z^\dag\right)^2\right)\bigg)+\mathcal{O}\left(\frac{\lambda}{J}\right)^3
\end{eqnarray}
(see \cite{Prosen2012} for a generalization to the case of asymmetric driving and \cite{Prosen2014} for a review). Here $Z$ is a matrix product operator
\begin{equation} \label{MPO}
Z=\sum_{\{s_1,\dots,s_N\}\in\{0,+,-\}^N}\langle L|\prod_{j=1}^nA_{s_j}|R\rangle\bigotimes_{j=1}^n\sigma_j^{s_j}
\end{equation}
with tridiagonal matrices $A_{s_j}$ on the auxiliary Hilbert space ${\cal H}$ spanned by the orthonormal basis $\{|L\rangle,|R\rangle,|1\rangle,|2\rangle,\dots,|\lfloor\frac{n}{2}\rfloor\rangle\}$:
\begin{eqnarray}
A_0 & = & |L\rangle\langle L|+|R\rangle\langle R|+\sum_{k=1}^{\lfloor \frac{n}{2}\rfloor}\cos(\eta k)|k\rangle\langle k|, \nonumber \\
A_+ & = & |1\rangle\langle R|-\sum_{k=1}^{\lfloor \frac{n}{2}\rfloor}\sin(\eta k)|k+1\rangle\langle k|, \nonumber  \\
A_- & = & |L\rangle\langle 1|+\sum_{k=1}^{\lfloor \frac{n}{2}\rfloor}\sin(\eta(k+1))|k\rangle\langle k+1|, \label{Apert}
\end{eqnarray}
and $\eta=\arccos\Delta\in\mathbbm{R}\cup i\mathbbm{R}$. The expansion \eqref{NESS.pert} holds as soon as the zeroth order is larger than the first order in $\frac{\lambda}{J}$. Estimating the magnitude of each order with its Hilbert-Schmidt norm ($||O||_{\textnormal{HS}}=\sqrt{\textnormal{Tr}(OO^\dag)}$), the validity condition for \eqref{NESS.pert} reads

\begin{equation} \label{val}
\frac{\lambda}{J}<\frac{\sqrt{2^{n+1}}}{\mu}||Z||_\textnormal{HS}^{-1}.
\end{equation}

If $\mu=1$ the non-perturbative NESS at any order of $\frac{\lambda}{J}$ \cite{Prosen2011b,Prosen2013b} is

\begin{equation} \label{MPO2}
\rho_\infty=\frac{SS^\dag}{\textnormal{Tr}(SS^\dag)}, \quad S=\sum_{\substack{\{s_1,\dots,s_n\}\\\in\{0,+,-\}^n}}\langle 0|\prod_{j=1}^nB_{s_j}|0\rangle\bigotimes_{j=1}^n\sigma_j^{s_j},
\end{equation}
with the matrix product operator $S$ and tridiagonal matrices $B_{s_j}$ on the auxiliary Hilbert space spanned by the orthonormal basis $\{|0\rangle,|1\rangle,|2\rangle,\dots,|\lfloor\frac{n}{2}\rfloor\rangle\}$

\begin{eqnarray}
B_0 & = & \sum_{k=0}^{\lfloor\frac{n}{2}\rfloor}\sin(\eta(s-k))|k\rangle\langle k|, \nonumber \\
B_+ & = & \sum_{k=0}^{\lfloor\frac{n}{2}\rfloor}\sin(\eta(k-2s))|k\rangle\langle k+1|, \nonumber \\
B_- & = & \sum_{k=0}^{\lfloor\frac{n}{2}\rfloor}\sin(\eta(k+1))|k+1\rangle\langle k|, \label{Bnonpert}
\end{eqnarray}
and with $s$ given by $\cot(s\eta)=\frac{\epsilon}{4i\sin\eta}$.

\section{Quantum estimation theory}

In this section, we discuss some fundamental aspects of quantum metrology \cite{Helstrom1976,Holevo1982,Braunstein1996,Paris2009}, relevant in our study. We are interested in estimating one of the parameters $x$ of the NESS, e.g. $\{J,\Delta,\lambda,\mu\}$. We can express the change of the NESS with respect to $x$ by the following equation

\begin{equation}
\frac{d}{dx}\rho_\infty=\frac{1}{2}\left(L_x\rho_\infty+\rho_\infty L_x\right),
\end{equation}
where

\begin{equation} \label{SLD}
L_x=2\int_0^\infty ds \, e^{-s\rho_{\infty}}\frac{d\rho_\infty}{dx}e^{-s\rho_{\infty}}
\end{equation}
is the symmetric logarithmic derivative which can in general depend on the parameter $x$ to be estimated. Quantum estimation theory gives the best achievable sensitivities, namely estimator variances, through the quantum Cram\'er-Rao bound \cite{Helstrom1976,Holevo1982,Braunstein1996,Paris2009}

\begin{equation} \label{CRB}
\delta^2x\geqslant\frac{1}{F_x},
\end{equation}
where

\begin{eqnarray}
F_x & = & \textnormal{Tr}\left(L_x^2\rho_\infty\right)=\textnormal{Tr}\left(L_x\frac{d\rho_\infty}{dx}\right)= \nonumber \\
& = & 2\int_0^\infty ds\,\textnormal{Tr}\left[\left(\frac{d\rho_\infty}{dx}e^{-s\rho_{\infty}}\right)^2\right] \label{Fisher}
\end{eqnarray}
is called Fisher information. The quantum Cram\'er-Rao bound \eqref{CRB} can be saturated by a projective measurement onto the eigenstates of the symmetric logarithmic derivative. This measurement may however depend on the parameter to be estimated, and thus might not be practically relevant. On the one hand, it is an open problem to find practical and optimal estiomations, on the other hand Fisher information itself provides the theoretical bound of the estimation sensitivity.
The best relative error are thus

\begin{equation} \label{relative}
\min_\textnormal{all estimations}\frac{\delta x}{x}=\frac{1}{x\sqrt{F_x}}.
\end{equation}

In the multiparameter estimation the inverse of covariance matrix of any estimation is bounded from below by the matrix $[F_{x,x'}]_{x,x'}$ with $F_{x,x'}=\frac{1}{2}\textnormal{Tr}\left(\rho_\infty\{L_x,L_{x'}\}\right)$. Thus, $1/F_x$ is still the best variance for the estimation of $x$, and the covariances depend on $F_{x,x'}$. Nevertheless, the matrix bound is not saturable in general because the optimal estimation of different parameters do not commute in general, and thus cannot be simultaneously performed. A tight multiparameter bound is still an open problem.

We compute the Fisher information \eqref{Fisher} for the non-perturbative NESS ($\mu=1$) numerically, and its leading order for the perturbative NESS \eqref{NESS.pert} analytically. At the lowest order in $\frac{\lambda}{J}$, the Fisher information is $F_x= F_x^{(0)}+\mathcal{O}\left(\frac{\lambda}{J}\right)^{\nu}$, where $\nu=2$ if $x\equiv \lambda$ and $\nu=4$  otherwise, and

\begin{eqnarray}
F_x^{(0)} & = & 2\int_0^\infty ds\,e^{-\frac{2s}{2^n}}\textnormal{Tr}\left[\left(\frac{d}{dx}\frac{i\lambda\mu}{J \, 2^{n+1}}(Z-Z^\dag)\right)^2\right]= \nonumber \\
& = & \frac{1}{2^{n+1}}\left|\left|\frac{d}{dx}\frac{\lambda\mu Z}{J}\right|\right|_\textnormal{HS}^2. \label{F0}
\end{eqnarray}
Note that the contributions of order $\left(\frac{\lambda}{J}\right)^{\nu-1}$ vanish because they are traces of real, antihermitian matrices, as can be realized plugging \eqref{NESS.pert} into \eqref{Fisher}. Only the parameter $\Delta$ enters non-trivially in the operator $Z$, whereas the others enter as multiplicative constants. Henceforth, we
consider the case of parameter $\Delta$ separately, provided the latter is independent from the other parameters.

\subsection{Computation of $F_x^{(0)}$ with $x\not\equiv\Delta$}

In this subsection we explicitly compute $F_x^{(0)}$ for $x$ different from $\Delta$. Plugging \eqref{MPO} and \eqref{Apert} into \eqref{F0}, we get

\begin{widetext}
\begin{eqnarray}
F_x^{(0)} & = & \frac{1}{2^{n+1}}\left(\frac{d}{dx}\frac{\lambda\mu}{J}\right)^2\sum_{\substack{\{s_1,\dots,s_n\},\{s'_1,\dots,s'_n\}\\\in\{0,+,-\}^n}}\langle L|\prod_{j=1}^nA_{s_j}|R\rangle\langle L|\prod_{j=1}^nA_{s'_j}|R\rangle \, \textnormal{Tr}\left[\bigotimes_{j=1}^n\sigma_j^{s_j}\left(\sigma_j^{s'_j}\right)^\dag\right]= \nonumber \\
& = & \frac{1}{2}\left(\frac{d}{dx}\frac{\lambda\mu}{J}\right)^2\langle L|\otimes\langle L|\left(A_0\otimes A_0+\frac{1}{2}A_+\otimes A_++\frac{1}{2}A_-\otimes A_-\right)^n|R\rangle\otimes|R\rangle.
\end{eqnarray}
\end{widetext}
Since the subspace $\{|k\rangle\otimes|k\rangle\}_{k=L,R,1,2,\dots,\lfloor\frac{N}{2}\rfloor}$ of the auxiliary space is preserved by operators $A_{s_j}\otimes A_{s_j}$, we apply the mapping $|k\rangle\otimes|k\rangle\to|k\rangle$, and finally obtain

\begin{equation} \label{F0x}
F_x^{(0)}=\left(\frac{d}{dx}\frac{\lambda\mu}{J}\right)^2\frac{\langle L|T^n|R\rangle}{2}, \qquad \forall \, x \textnormal{ except } \Delta,
\end{equation}

where the transfer matrix $T$ is the following

\begin{eqnarray}
T & = & |L\rangle\langle L|+|R\rangle\langle R|+\frac{|L\rangle\langle 1|+|1\rangle\langle R|}{2} \nonumber \\
&& +\sum_{k=1}^{\lfloor\frac{n}{2}\rfloor}\Bigg(\cos^2(\eta k)|k\rangle\langle k|+\frac{\sin^2(\eta k)}{2}|k+1\rangle\langle k| \nonumber \\
&& +\frac{\sin^2(\eta (k+1))}{2}|k\rangle\langle k+1|\Bigg). \label{transfer}
\end{eqnarray}

The smallest order Fisher information $F_x^{(0)}$ is then expressed in terms of the exponential of the transfer matrix $T$. This form enables the computation of $F_x^{(0)}$ using the combinatorics of $n$-step paths going from $|R\rangle$ to $|L\rangle$ by means of climbs and descents through intermediate states $\{|k\rangle\}_{k\geqslant1}$. The amplitude of each step is the corresponding matrix element of the transfer matrix $T$.

The optimal estimator of any parameter $x$ except $\Delta$, attaining the quantum Cram\'er-Rao bound \eqref{CRB} at the lowest order in $\frac{\lambda}{J}$, is devised by a measurement 
$\textnormal{Tr}\left(\zeta\rho_\infty\right)$ of an observable given by hermitian operator $\zeta\equiv~i(Z-Z^\dag)$. The experimental input is $\zeta_m=\frac{1}{m}\sum_{j=1}^mz_j$, where $\{z_j\}_{j=1,\dots,m}$ are $m$ measurement outcomes of $\zeta$. Each value $z_j$ is sampled from the probability distribution of measuring such value from a system in the state $\rho_\infty$ \footnote{Given an eigenvalue $z$ of the observable $\zeta$, corresponding to the eigenvector $|z\rangle$, a measure of $\zeta$ with a system in the state $\rho_\infty$ provides the value $z$ with probability $\langle z|\rho_\infty|z\rangle$.}. The average and the variance of $\zeta_m$ with respect to this probability distribution are

\begin{eqnarray}
\mathbb{E}[\zeta_m] & = & \textnormal{Tr}(\zeta\rho_\infty), \nonumber \\
\textnormal{Var}\zeta_m & \equiv & \mathbb{E}\left[\left(\zeta_m-\textnormal{Tr}(\zeta\rho_\infty)\right)^2\right] \nonumber \\
& = & \frac{1}{m}\left(\textnormal{Tr}(\zeta^2\rho_\infty)-(\textnormal{Tr}\zeta\rho_\infty)^2\right), \label{var.zeta}
\end{eqnarray}
For large $m$, $\zeta_m\to\textnormal{Tr}(\zeta\rho_\infty)$ is the statistical average of experimental outcomes which converges to the expectation of the operator $\zeta$. Inverting the relation $\zeta_m=\textnormal{Tr}(\zeta\rho_\infty)$, we estimate $x$ with sensitivity at the lowest order in $\frac{\lambda}{J}$

\begin{equation}
\delta^2x=\frac{\textnormal{Var}\zeta_m}{\left(\frac{d}{dx}\textnormal{Tr}(\zeta\rho_\infty)\right)^2}=\frac{1}{mF_x^{(0)}}, \quad \forall \, x \textnormal{ except } \Delta.
\end{equation}

\subsection{Computation of $F_\Delta^{(0)}$}

We now derive an explicitly formula for $F_\Delta^{(0)}$. With the expressions \eqref{MPO} and \eqref{Apert}, equation \eqref{F0} for $x\equiv\Delta$ becomes

\begin{widetext}
\begin{equation}
F_\Delta^{(0)}=\frac{1}{2^{n+1}}\left(\frac{\lambda\mu}{J}\right)^2\sum_{\substack{\{s_1,\dots,s_n\},\{s'_1,\dots,s'_n\}\\\in\{0,+,-\}^n}}\langle L|\frac{d}{d\Delta}\left(\prod_{j=1}^nA_{s_j}\right)|R\rangle\langle L|\frac{d}{d\Delta}\left(\prod_{j=1}^nA_{s'_j}\right)|R\rangle \, \textnormal{Tr}\left[\bigotimes_{j=1}^n\sigma_j^{s_j}\left(\sigma_j^{s'_j}\right)^\dag\right].
\end{equation}
\end{widetext}
After some algebra, and using the above mapping $|k\rangle\otimes|k\rangle\to|k\rangle$ and $\frac{d}{d\Delta}=\frac{d\mu}{d\Delta}\frac{d}{d\mu}=-\frac{1}{\sqrt{1-\Delta^2}}\frac{d}{d\mu}$, we get

\begin{equation} \label{F0Delta}
F_\Delta^{(0)}=\frac{\lambda^2\mu^2}{2J^2(1-\Delta^2)}\langle L|\left(\sum_{k=1}^nT^{k-1}DT^{n-k}+\frac{1}{4}\frac{d^2T^n}{d\eta^2}\right)|R\rangle,
\end{equation}
with the vertex matrix

\begin{eqnarray}
D & = & \sum_{k=1}^{\lfloor \frac{n}{2}\rfloor}\Big(\textnormal{sign}(1-\Delta^2)\frac{k^2}{2}|k\rangle\langle k|+\frac{k^2}{4}|k+1\rangle\langle k| \nonumber \\
&& +\frac{(k+1)^2}{4}|k\rangle\langle k+1|\Big), \label{vertex}
\end{eqnarray}

In analogy to $F_x^{(0)}$, also the Fisher information $F_\Delta^{(0)}$ is expressed in terms of the transfer matrix $T$ with the insertion of the defect matrix $D$. Therefore, we can still use the combinatorial picture of paths going from $|R\rangle$ to $|L\rangle$, climbing and descending via intermediate states $\{|k\rangle\}_{k\geqslant1}$, where one of the $n$-th steps is ruled by the matrix element of $D$.

\section{Scaling of the Fisher information and the relative error}

In this section, we study the scaling with the particle number $n$ of the Fisher information and of the relative error of parameter estimations in different interaction regimes. A first general remark is that the validity condition \eqref{val} of the perturbation expansion \eqref{NESS.pert}, together with the lowest order Fisher information \eqref{F0}, implies that the best relative error of $x\not\equiv\Delta$ is pretty large for any value of $F_x^{(0)}$: $\frac{1}{x\sqrt{F_x^{(0)}}}>1$. This is not the case for the anisotropy $\Delta$, as we discuss in the following. Moreover, $F_x^{(0)}$ for any $x$ show peculiar scalings in the three phases: isotropic interaction $|\Delta|=1$, easy-plane interactions $|\Delta|<1$, and easy-axis interactions $|\Delta|>1$.

\subsection{Isotropic limit $|\Delta|=1$}

We now show in the isotropic limit $\Delta\to1$ and $\eta\to0$ of the perturbative NESS how the computation of the low orders in $\eta$ of the leading contribution for small $F_x^{(0)}$ is equivalent to the combinatorics of a random walk between the states $|R\rangle$ and $|L\rangle$ with transition amplitudes given by the transfer matrix. First, we expand the sinusoidal functions in the transfer matrix \eqref{transfer} for small arguments $k\eta\ll1$, $\forall\,k=1,\dots,\lfloor\frac{n}{2}\rfloor$, thus $\eta\ll\frac{2}{n}$: $T=T_0+\sum_{l=1}^\infty\eta^{2l}T_{2l}$ with

\begin{eqnarray}
T_0 & \!\! = & \!\! |L\rangle\langle L|+|R\rangle\langle R|+|1\rangle\langle 1|+\frac{|1\rangle\langle R|+|L\rangle\langle 1|}{2} \nonumber \\
T_{2l>0} & \!\! = & \!\! \sum_{k=1}^{\lfloor\frac{n}{2}\rfloor}\Bigg((-)^l\frac{(2k)^{2l}}{2(2l)!}|k\rangle\langle k|+(-)^{l-1}\frac{(2k)^{2l}}{4(2l)!}|k+1\rangle\langle k| \nonumber \\
&& +(-)^{l-1}\frac{(2(k+1))^{2l}}{4(2l)!}|k\rangle\langle k+1|\Bigg).
\end{eqnarray}
Expanding the exponential $T^n$, the lowest orders in $\eta$ of \eqref{F0x} and \eqref{F0Delta} are sums of matrix elements between the states $|R\rangle$ and $|L\rangle$ of products with many transfer matrices $T_0$ and a few vertices $T_{2l>0}$ and $D$. Finally, we obtain

\begin{eqnarray}
\!\! \langle L|T^n|R\rangle & \!\! =  & \!\! \frac{n(n-1)}{8}-\frac{n(n-1)(n-2)}{24}\bigg(\eta^2-\frac{\eta^4}{6}(3n-7) \nonumber \\
\!\! & \!\! & \!\! +\frac{\eta^6}{180}(989+3n(36n-217)\bigg)+\mathcal{O}(\eta^8), \\
\!\! F_\Delta^{(0)} & \!\! = & \!\! \frac{\lambda^2}{96 \, J^2} \, \mu^2n(n-1)(n-2)\bigg(3n-7 \nonumber \\
\!\! & \!\! & \!\! -\frac{\eta^2}{30}(n-3)(261n-799)\bigg)+\mathcal{O}\left(\frac{\lambda^2}{J^2}\eta^4\right).
\end{eqnarray}

We already commented on the relative error for $x\not\equiv\Delta$, that is larger than one. The best relative error of $\Delta$ in the isotropic limit is

\begin{equation} \label{rel.sens.isot}
\frac{1}{\Delta F_\Delta^{(0)}}>\mathcal{O}\left(\frac{1}{n^2}\right),
\end{equation}
where we have used the validity condition \eqref{val} for small coupling $\frac{\lambda}{J}$, $\langle L|T^n|R\rangle=\mathcal{O}(n^2)$ and $F_\Delta^{(0)}=\mathcal{O}(n^4)$. Equation \eqref{rel.sens.isot} implies that the relative error of the anisotropy $\Delta$ can decrease faster than the shot-noise-limit with increasing size $n$, e.g. $1/n^{1+\alpha}$ with $\alpha\in(0,1)$.

\subsection{Anisotropic regime: easy-plane interactions $|\Delta|<1$}

The scaling of Fisher information changes in the deep anisotropic regime. The leading orders in $\lambda$ of $F_x^{(0)}$ for infinitely large $n$ can be explicitly computed from the Jordan block of $T$ corresponding to the largest eigenvalue: see appendix \ref{app} for the Jordan decomposition of the transfer matrix. We consider two complementary cases: a dense subinterval of $\Delta\in(-1,1)$, namely $\eta=\frac{q\pi}{p}$ with coprime integers $p,q$ without loss of generality, and the case of irrational $\frac{\eta}{\pi}$. These cases exhibit different behaviours for $F^{(0)}_\Delta$.

\begin{figure}[htbp]
\centering
\includegraphics[width=\columnwidth]{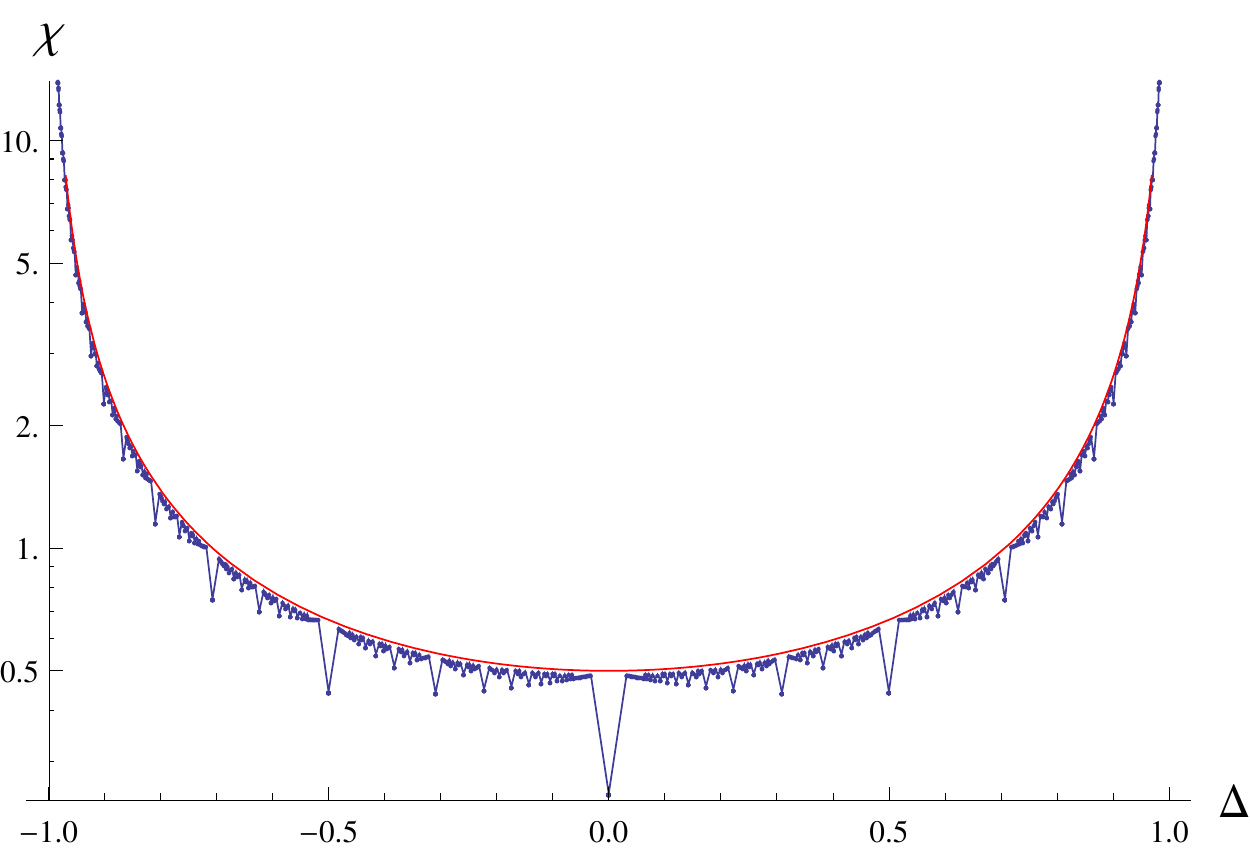}
\caption{(Colours online) Semi-log plots of the coefficients $\chi$, determining the size scaling of Fisher information of NESS, as functions of $\Delta$ with $|\Delta|<1$ for rational (dots joined by the blue line) $\frac{\eta}{\pi}$ and irrational (red line) $\frac{\eta}{\pi}$ for large $n$. The two lines perfectly overlap except for some spikes for small denominators of $\frac{\eta}{\pi}$.}
\label{chi.fig}
\end{figure}

In the first case $\eta=\frac{q\pi}{p}$, the transitions $\big\langle|p|\pm1\big|T\big||p|\big\rangle$ vanish. Therefore, we can consider the restriction $T^{(d)}$ of the transfer matrix \eqref{transfer} in the basis $\{|L\rangle,|R\rangle,|1\rangle,|2\rangle,\dots,|d\rangle\}$ with $d=|p|-1$ for the computation of the matrix power in the expression \eqref{F0x} of $F_x^{(0)}$ with $x\not\equiv\Delta$. For irrational $\frac{\eta}{\pi}$, none of the transitions $\big\langle|p|\pm1\big|T\big||p|\big\rangle$ vanish and we need the full transfer matrix, thus with $d=\lfloor\frac{n}{2}\rfloor$. In both cases, the matrix 
$T^{(d)}$ is transformed into its Jordan canonical form via the similarity $T_J^{(d)}=(V^{(d)})^{-1} T^{(d)} V^{(d)}$, see appendix \ref{app} for details. The largest eigenvalue of $T^{(d)}$ is $1$ corresponding to the eigenvector $|L\rangle$ and a defective eigenvector, and the other eigenvalues $\{\tau_j\}_{j=1,\dots,d}$ satisfy $1>|\tau_1|\geqslant\dots\geqslant|\tau_d|$. One analytically computes

\begin{equation}
\langle L|T^n|R\rangle=\langle L|V^{(d)}(T^{(d)}_J)^n(V^{(d)})^{-1}|R\rangle=\chi n+\chi_1,
\end{equation}
with

\begin{eqnarray}
\chi & = & \langle L|V^{(d)}|L\rangle\langle R|(V^{(d)})^{-1}|R\rangle=\frac{d}{2(d+1)}\cdot\frac{1}{1-\Delta^2}, \nonumber \\
\chi_1 & = & \langle L|(V^{(d)})^{-1}|R\rangle.
\end{eqnarray}
These equations show that the Fisher information $F_x^{(0)}$ with $x\not\equiv\Delta$ is linear in the particle number. Figure \ref{chi.fig} shows the coefficient $\chi$ for rational and irrational $\frac{\eta}{\pi}$ in the leading order for large $n$: note that there is no qualitative difference between the two cases. Remember however that despite the linear scaling of the Fisher information, the relative error of $x\not\equiv\Delta$ is larger than $1$ within the perturbation analysis \eqref{NESS.pert} and \eqref{val}.

The computation of $F_\Delta^{(0)}$ at the lowest order for $n\gg1$ is more involved. The second term of \eqref{F0Delta} is

\begin{equation}
\frac{d^2}{d\eta^2}\langle L|T^n|R\rangle=\frac{d^2\chi n}{d\eta^2}+\mathcal{O}(1)=\frac{d}{d+1}\cdot\frac{2\Delta^2+1}{(1-\Delta^2)^2} \, n+\mathcal{O}(1),
\end{equation}
with $\mathcal{O}(1)$ being constants in $n$. The other contribution of \eqref{F0Delta} differentiates between rational and irrational $\frac{\eta}{\pi}$. If $\eta=\frac{q\pi}{p}$, the transition $\big\langle|p|-1\big|T\big||p|\big\rangle=0$ implies that the transfer matrices on the left of the vertex matrix $D$ can be restricted to $T^{(d)}$ with $d=|p|-1$, as well as for the transfer matrices on the right of $D$ except the one adjacent to $D$. The reason for this exception is that the transition $\big\langle|p|-1\big|D\big||p|\big\rangle$ does not vanish, and thus the next $T$ matrix on the right of $D$ can lead the transition $\big||p|-1\big\rangle\to\big||p|\big\rangle$. We single out the latter contribution in equation \eqref{xi.rat}. We again use the Jordan decomposition \eqref{powerT} and $d=|p|-1$:

\begin{widetext}
\begin{eqnarray}
&& \sum_{k=1}^n\langle L|T^{k-1}DT^{n-k}|R\rangle=\sum_{k=1}^n\sum_{j=1}^d\sum_{l=1}^{d+1}\langle L|(T^{(d)})^{k-1}|j\rangle\langle j|D|l\rangle\langle l|(T^{(d)})^{n-k}|R\rangle= \nonumber \\
&& =\sum_{k=1}^n\sum_{j,l=1}^d\langle L|V^{(d)}(T^{(d)}_J)^{k-1}(V^{(d)})^{-1}|j\rangle\langle j|D|l\rangle\langle l|V^{(d)}(T^{(d)}_J)^{n-k}(V^{(d)})^{-1}|R\rangle \nonumber \\
&& +\sum_{k=1}^n\langle L|V^{(d)}(T^{(d)}_J)^{k-1}(V^{(d)})^{-1}|d\rangle\langle d|D|d+1\rangle\langle d+1|T|d\rangle\langle d|V^{(d)}(T^{(d)}_J)^{n-k-1}(V^{(d)})^{-1}|R\rangle= \nonumber \\
&& =\frac{d}{2(d+1)}\cdot\frac{n}{1-\Delta^2}\left(\sum_{j,l=1}^d\langle j|D|l\rangle\langle L|V^{-1}|j\rangle\bar\psi_l+\frac{(d+1)^2}{4}\cdot\frac{\sin((d\eta)}{2}\langle L|V^{-1}|d\rangle\bar\psi_d\right)+\mathcal{O}(1)\equiv\xi_1 n+\mathcal{O}(1). \label{xi.rat}
\end{eqnarray}
\end{widetext}
For irrational $\frac{\eta}{\pi}$, the transfer matrix in any position always equals $T^{(d)}$ with $d=\lfloor\frac{n}{2}\rfloor$. In this case, the first term of \eqref{F0Delta} becomes

\begin{widetext}
\begin{eqnarray}
&& \sum_{k=1}^n\langle L|T^{k-1}DT^{n-k}|R\rangle=\sum_{k=1}^n\sum_{j,l=1}^d\langle L|(T^{(d)})^{k-1}|j\rangle\langle j|D|l\rangle\langle l|(T^{(d)})^{n-k}|R\rangle= \nonumber \\
&& =\sum_{k=1}^n\sum_{j,l=1}^d\langle L|V^{(d)}(T^{(d)}_J)^{k-1}(V^{(d)})^{-1}|j\rangle\langle j|D|l\rangle\langle l|V^{(d)}(T^{(d)}_J)^{n-k}(V^{(d)})^{-1}|R\rangle= \nonumber \\
&& =\frac{d}{2(d+1)}\cdot\frac{n}{1-\Delta^2}\sum_{j,l=1}^d\langle j|D|l\rangle\langle L|V^{-1}|j\rangle\bar\psi_l+\mathcal{O}(1)\equiv\xi_1 n+\mathcal{O}(1). \label{xi.irrat}
\end{eqnarray}
\end{widetext}
Finally, we obtain at the leading order for large $n$

\begin{eqnarray}
F_\Delta^{(0)} & = & \frac{\lambda^2\mu^2}{J^2}\xi n+\mathcal{O}(1) \qquad \textnormal{with} \nonumber \\
\xi & = & \frac{1}{2(1-\Delta^2)}\left(\xi_1-\frac{1}{4}\frac{d^2\chi}{d\eta^2}\right)
\end{eqnarray}

\begin{widetext}

\begin{figure}[htbp]
\centering
\includegraphics[width=0.49\columnwidth]{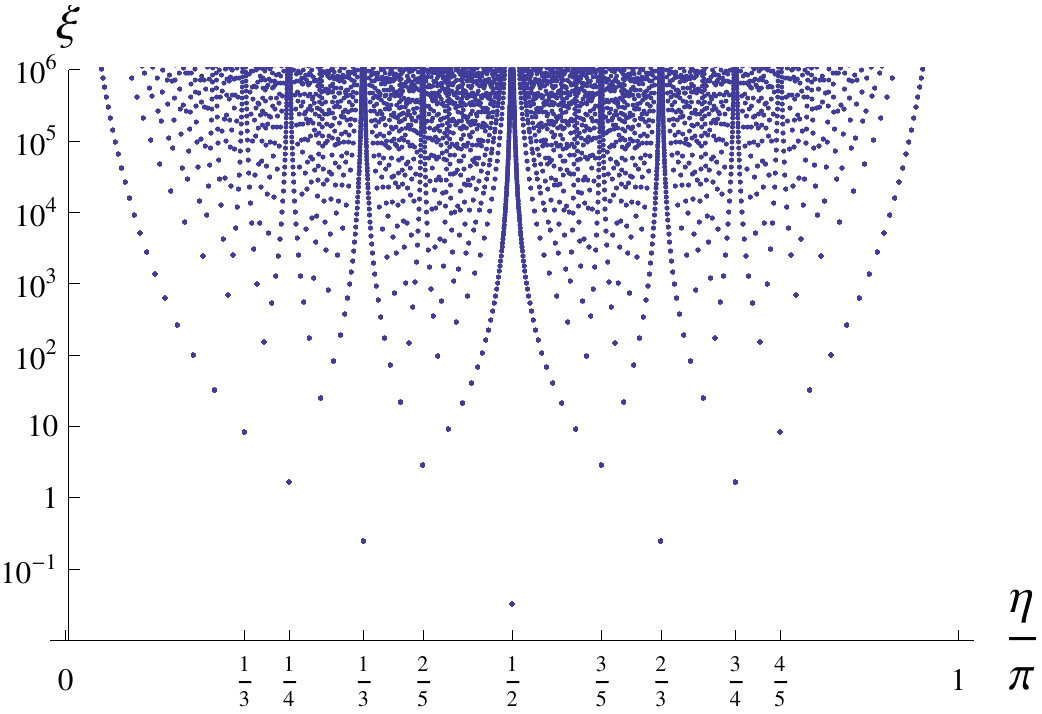}
\includegraphics[width=0.49\columnwidth]{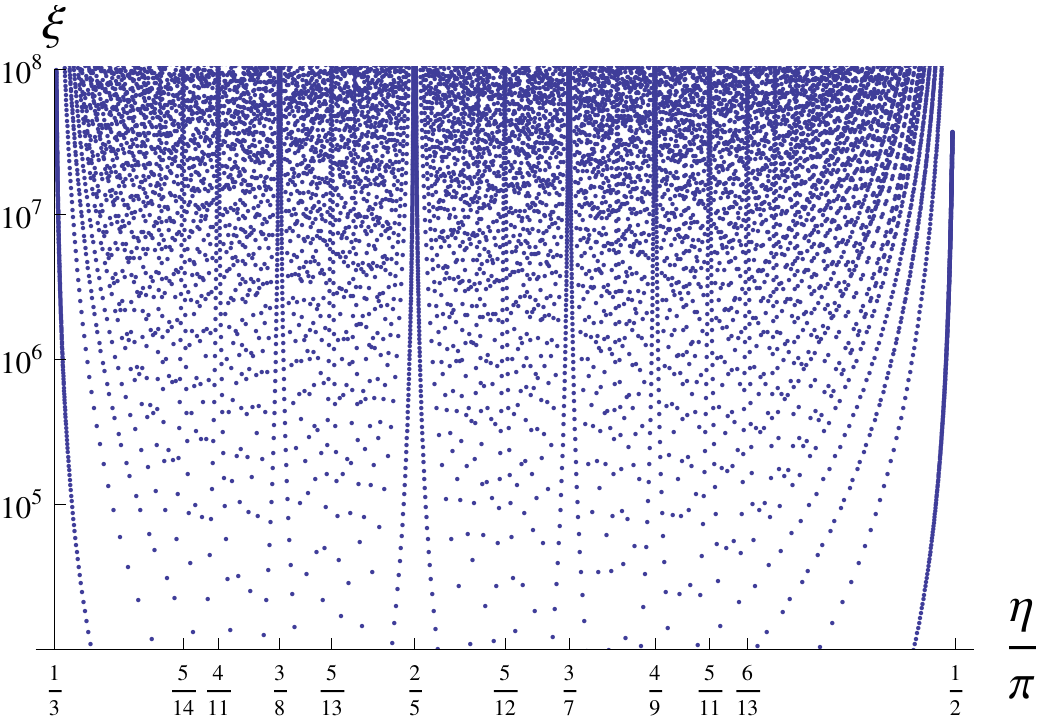}
\includegraphics[width=0.49\columnwidth]{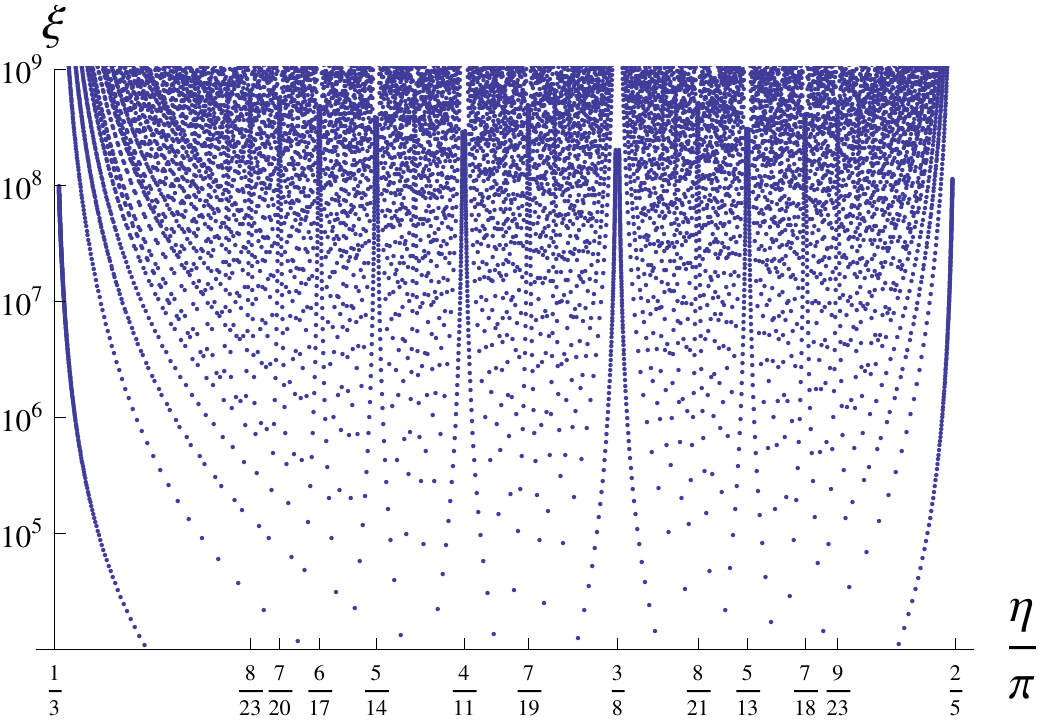}
\includegraphics[width=0.49\columnwidth]{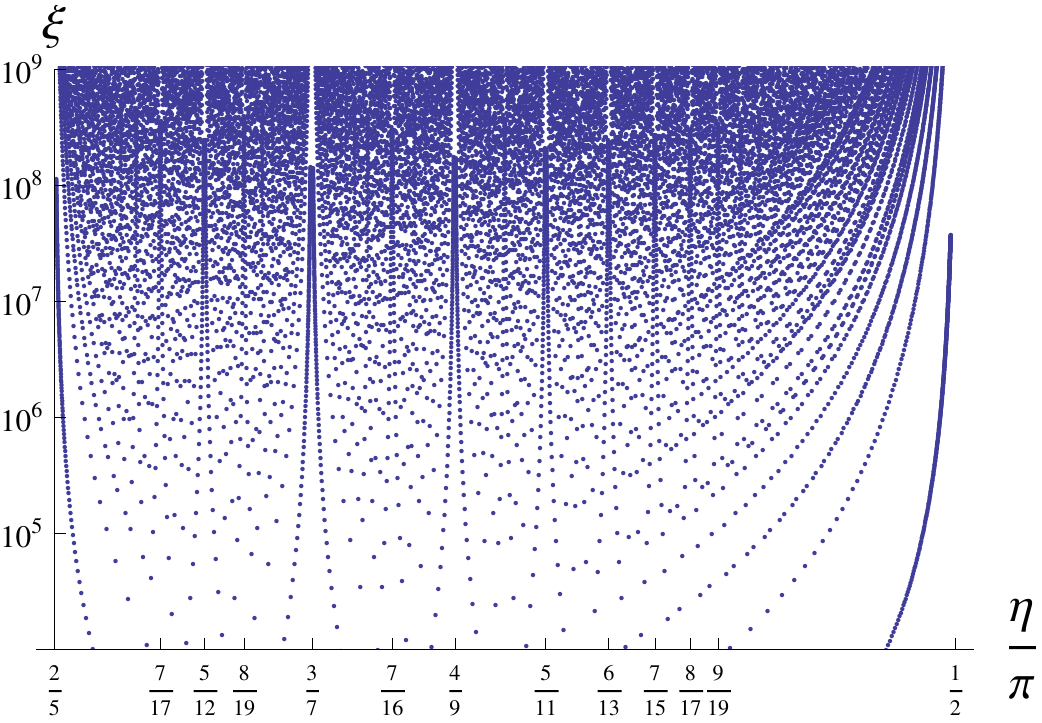}
\includegraphics[width=0.49\columnwidth]{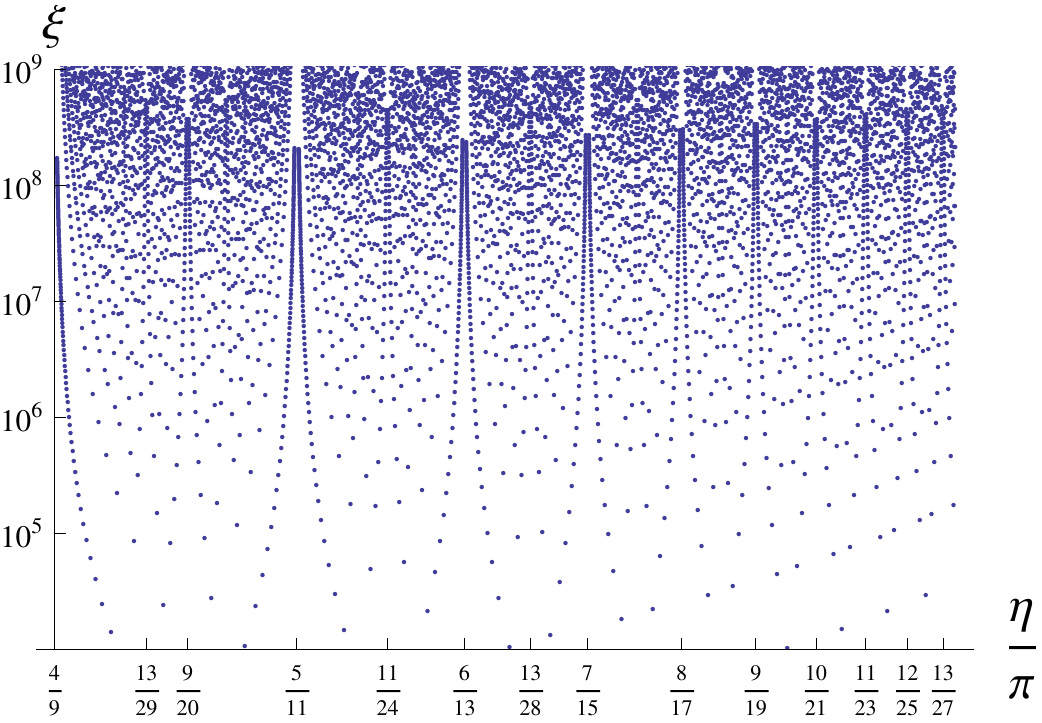}
\includegraphics[width=0.49\columnwidth]{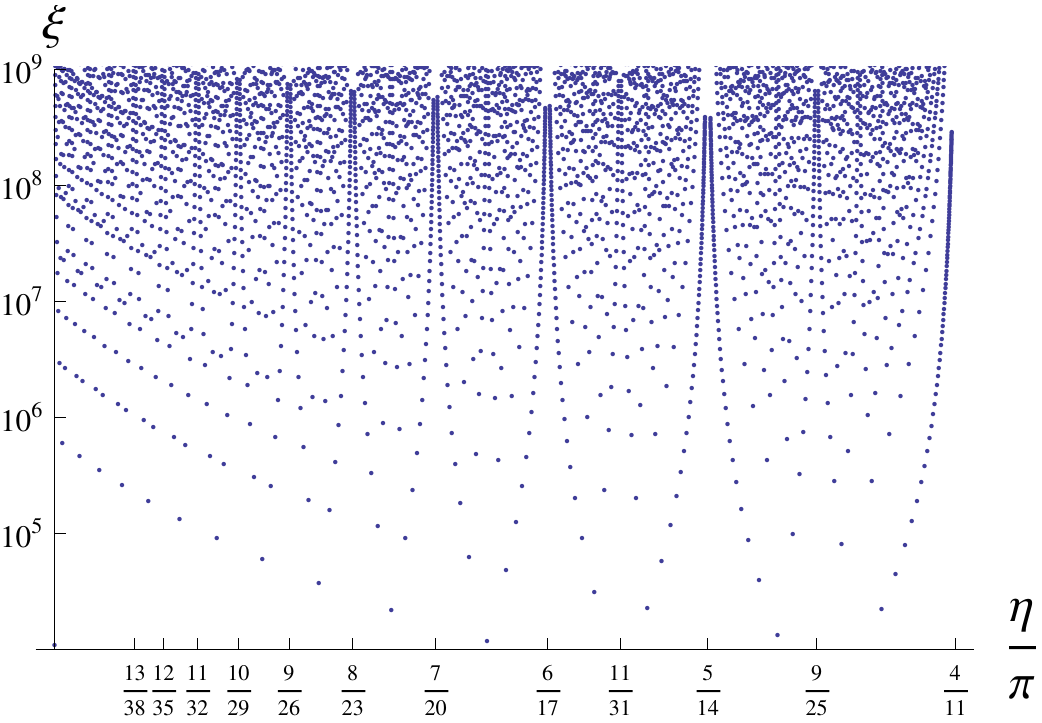}
\caption{Semi-log plots of the coefficient $\xi$ of the Fisher information with respect to the anisotropy as function of the anisotropy parameter $\frac{\eta}{\pi}\in[0,1]$, for rational $\frac{\eta}{\pi}$. We notice that closer zooms reveals additional cusp-like patterns, suggesting a fractal geometry. The blank spaces above some of the cusp-like patterns are due to the numerical accuracy: we generated rational $\frac{\eta}{\pi}=\frac{q}{p}$ with $p\in[2,1000]$ and $q\in[1,1000]$, but for larger values of $q$ and $p$ the blank spaces are filled. For larger values of $\xi$, the less evident cusp-like patterns become more evident.}
\label{xi.fig}
\end{figure}
\end{widetext}

\begin{widetext}

\begin{figure}[htbp]
\centering
\includegraphics[width=0.49\columnwidth]{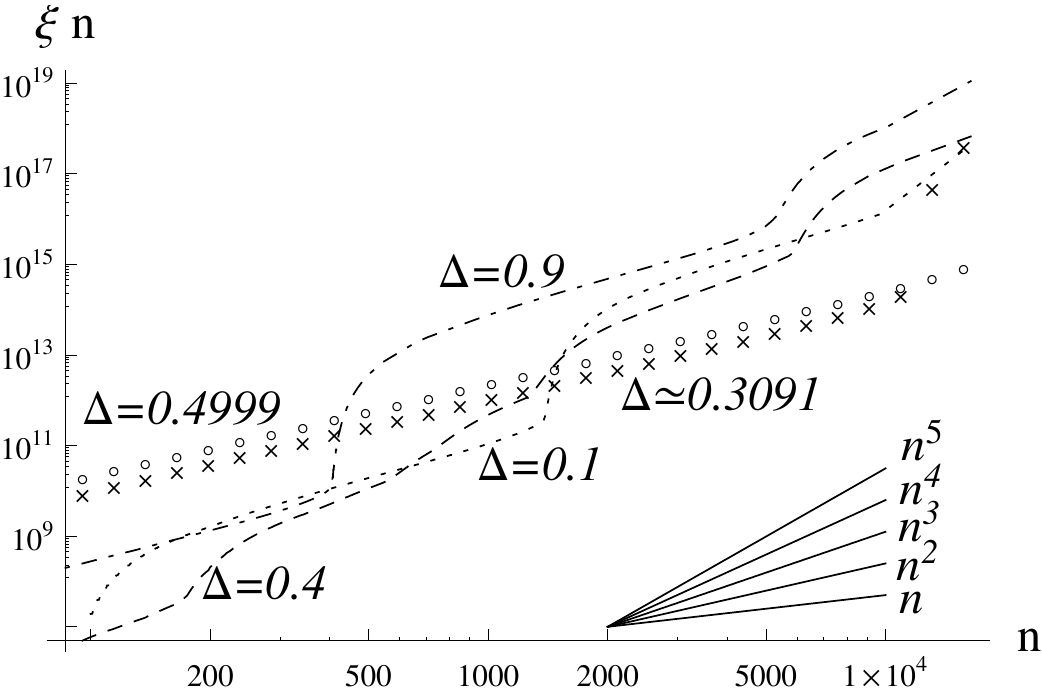}
\includegraphics[width=0.49\columnwidth]{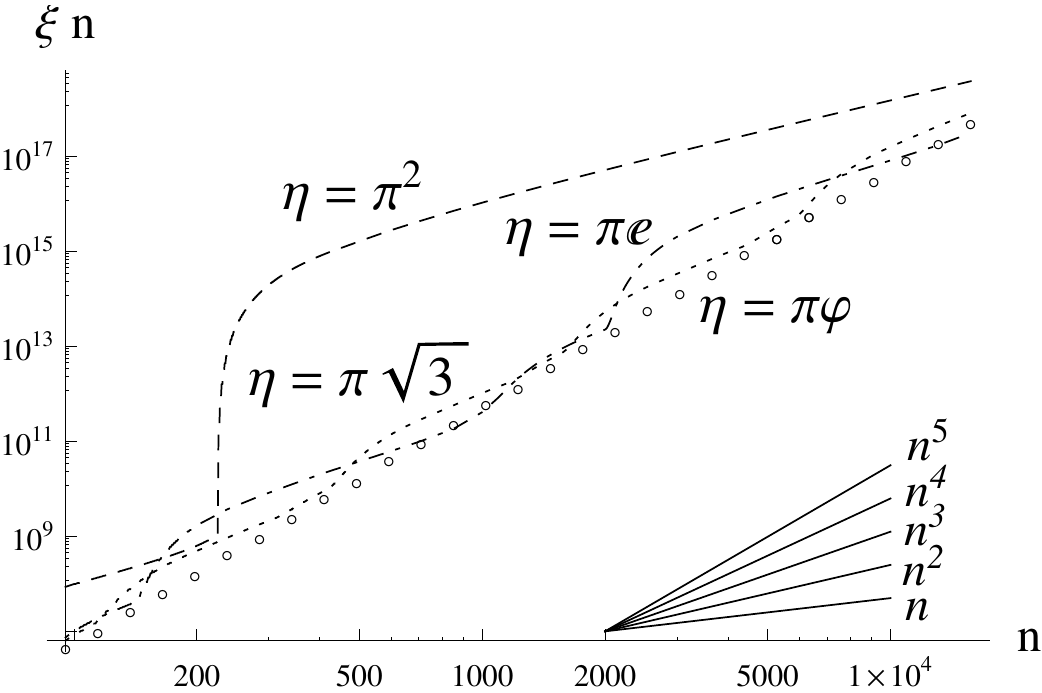}
\caption{Log-log plots of the coefficient $\xi n$, determining the size scaling of the Fisher information with respect to the anisotropy, as function of $n$ for irrational $\frac{\eta}{\pi}$. Left panel: $\Delta=0.1$ (dotted line), $\Delta=0.4$ (dashed line), and $\Delta=0.9$ (dotdashed line), $\Delta=0.4999$ (circles), $\Delta=\cos\left(\frac{2}{5}\pi\right)+10^{-4}\simeq0.390117$ (crosses). Right panel: $\eta=\pi\varphi$ with $\varphi=\frac{1+\sqrt{5}}{2}$ the golden ratio (circles), $\eta=\pi\sqrt{3}$ (dotted), $\eta=\pi^2$ (dashed), $\eta=\pi e$ (dotdashed). For a comparison we also plotted the slopes of power laws $n^1$, $n^2$, $n^3$, and $n^4$ (continuous lines).}
\label{xi-irr.fig}
\end{figure}
\end{widetext}

The coefficient $\xi$ is plotted in figure \ref{xi.fig}. It shows a very different behaviour from $\chi$, despite the fact that they are both no-where continuous functions of $\Delta$. While $\chi$ has a smooth envelope, $\xi$ is constant in $n$ but has a highly complex structure for rational $\frac{\eta}{\pi}$ with cusp-like patterns which can be appreciated by consecutive zooms suggesting a fractal geometry (see figure \ref{xi.fig}). Moreover, $\xi$ is unbounded for $|p|\to\infty$, indicating that the scaling of $F_\Delta^{(0)}$ is qualitatively different at irrational $\frac{\eta}{\pi}$. The differences between the coefficients $\chi$ and $\xi$, and therefore between \eqref{F0x} and \eqref{F0Delta}, are magnified for irrational $\frac{\eta}{\pi}$ since $\xi$ depends on $n$ as shown in \eqref{xi.irrat}: $\xi n$ plotted in figure \ref{xi-irr.fig} exhibits piecewise power scaling $\propto n^\alpha$ with $\alpha\geqslant2$ and an overall growth as fast as $\sim n^5$. Two remarks are in order. First, when $\frac{\eta}{\pi}$ is close to a rational number, like $\Delta=\cos\left(\frac{\pi}{3}\right)-10^{-4}=0.4999$ and $\Delta=\cos\left(\frac{2\pi}{7}\right)+10^{-4}\simeq0.390117$ in figure \ref{xi-irr.fig}, we observed the coefficient $\xi n$ grows like $n^2$ for a wide range of $n$ and then eventually increases the slope; this is a signature of the transition toward rational values of $\frac{\eta}{\pi}$ where $\xi$ is constant in $n$. Second, the oscillatory behaviour in figure \ref{xi-irr.fig} is less pronounced when $\frac{\eta}{\pi}$ is an algebraic irrational number, like the golden ratio $\varphi=\frac{1+\sqrt{5}}{2}$ and $\sqrt{3}$ in figure \ref{xi-irr.fig}.

The best relative error of $\Delta$ for easy-plane interactions is

\begin{equation} \label{rel.sens.easy.plane}
\frac{1}{\Delta F_\Delta^{(0)}}>\mathcal{O}\left(\frac{1}{\xi n}\right)=
\begin{cases}
\displaystyle \mathcal{O}\left(\frac{1}{\xi}\right) & \displaystyle \textnormal{for rational } \frac{\eta}{\pi} \\
\displaystyle \mathcal{O}\left(\frac{1}{n^4}\right) & \displaystyle \textnormal{for irrational } \frac{\eta}{\pi}
\end{cases},
\end{equation}
where we have used the validity condition \eqref{val} for small coupling $\frac{\lambda}{J}$, $\langle L|T^n|R\rangle=\mathcal{O}(n)$ and $\xi n=\mathcal{O}(n^5)$ for irrational $\frac{\eta}{\pi}$. Therefore, the relative error of the anisotropy $\Delta$ can be very small even though it does not decrease with $n$ for rational $\frac{\eta}{\pi}$, while can decrease faster than the shot-noise-limit with increasing size $n$, e.g. $\sim1/n^4$, for irrational $\frac{\eta}{\pi}$.

\subsection{Anisotropic regime: easy-axes interactions $|\Delta|>1$}

If $|\Delta|>1$ ($\eta\in i \, \mathbbm{R}$), $\langle L|T^n|R\rangle$ exhibits a superexponential scaling with $n$. This can be seen by noticing that $\langle L|T^n|R\rangle$ is the sum of positive terms each representing the transition from $|R\rangle$ to $|L\rangle$ through different paths. For a very comfortable estimation we consider $n$ even and only a single path with the first $\frac{n}{2}$ steps which increase the index of the auxiliary basis and the last $\frac{n}{2}$ steps which decrease the index:

\begin{equation}
\langle L|A_-^{\frac{n}{2}}A_+^{\frac{n}{2}}|R\rangle=\frac{1}{2^n}\prod_{k=1}^{\frac{n}{2}-1}\sin^2(k\eta)\sin^2((k+1)\eta).
\end{equation}
Using $\sinh^2y\geqslant y^2,\,\forall\,y\in\mathbbm{R}$, we find

\begin{equation}
\langle L|T^n|R\rangle>\frac{\eta^{2(n-2)}}{2^n}\left(\left(\frac{n}{2}\right)!\left(\frac{n}{2}-1\right)!\right)^2
\end{equation}
which grows faster than any exponential of $n$. A similar reasoning holds for odd $n$ and for $F_\Delta^{(0)}$.

\begin{figure}[htbp]
\centering
\includegraphics[width=\columnwidth]{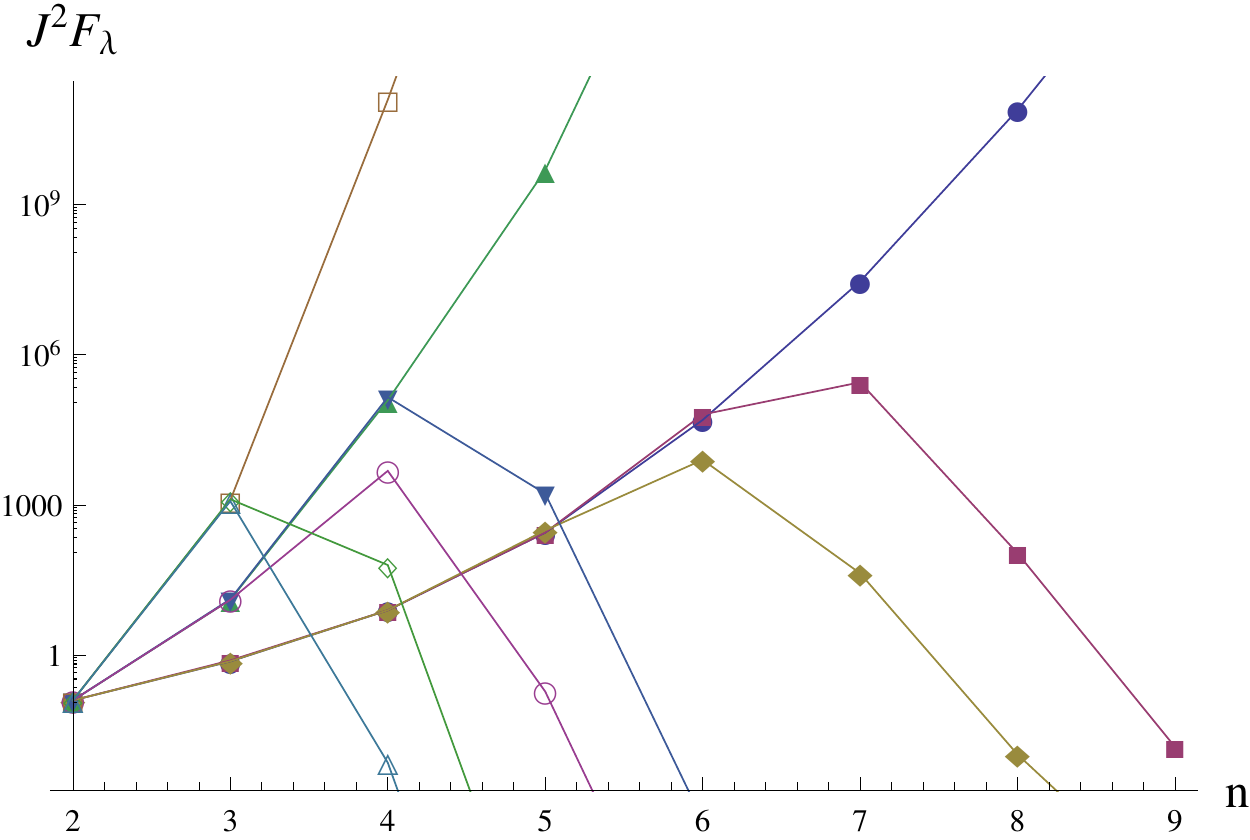}
\caption{(Color online) Semi-log plot of $J^2F_{\lambda}$ as function of $n$: $\left(\Delta,\frac{\lambda}{J}\right)=(2,0)$ (blue, full circles), $\left(\Delta,\frac{\lambda}{J}\right)=(2,10^{-3})$ (purple, full squares), $\left(\Delta,\frac{\lambda}{J}\right)=(2,10^{-2})$ (yellow, full diamonds), $\left(\Delta,\frac{\lambda}{J}\right)=(10,0)$ (green, full upward triangles), $\left(\Delta,\frac{\lambda}{J}\right)=(10,10^{-3})$ (blue, downward triangles), $\left(\Delta,\frac{\lambda}{J}\right)=(10,10^{-2})$ (purple, empty circles), $\left(\Delta,\frac{\lambda}{J}\right)=(10^2,0)$ (yellow, empty squares), $\left(\Delta,\frac{\lambda}{J}\right)=(10^2,10^{-3})$ (green, empty diamonds), $\left(\Delta,\frac{\lambda}{J}\right)=(10^2,10^{-2})$ (blue, empty upward triangle).}
\label{superexp}
\end{figure}

Such superexponential scaling is observed only perturbatively in $\lambda$, where the approximation holds if higher orders in $\lambda$ are small. From the non-perturbative solution at extreme driving $\mu=1$ \cite{Prosen2011b,Prosen2013b},
with NESS density operator \eqref{MPO2}, we have studied the behavior of $F_\lambda$ on $n$ numerically. We found that $F_\lambda$ grows superexponentially at small $n$ and then quickly decays to zero for small coupling $\lambda$, even though, as we have shown before, the leading order perturbative result keeps growing superexponentially. This can be understood as a crossover behavior from a highly correlated state, when the correlation length $\ell=\ell(\Delta)$ is of the order of the system size $\ell\approx n$ to a very simple, macroscopically almost separable kink between spin-up and spin-down ferromagnetic domains for $n\gg\ell$. In Fig. \ref{superexp}, we plotted $F_\lambda$ for $\Delta\in\{2,10,100\}$ and $\frac{\lambda}{J}\in\{0,10^{-2},10^{-3}\}$, where the cases $\frac{\lambda}{J}=0$ represent the leading orders of the perturbative expansions in $\frac{\lambda}{J}$. 

We have already noted that the relative error of the estimation of $\lambda$ is always pretty large, despite large values of the Fisher information $F_\lambda$. For $F_\Delta$ we only found small peaks, namely $F_\Delta\lesssim10$, and thus large relative error for the data in Fig. \ref{superexp}.

\section{Conclusions}

We have studied the NESS of an $XXZ$ spin chain with local dissipation at its ends as a probe for the estimation of several parameters, like spin interaction strength $J$, anisotropy $\Delta$, or dissipation parameters $\lambda$ and $\mu$. We quantified the efficiency of the above parameter estimations with the Fisher information, and studied the scaling with the number of spins. We found superlinear scaling for small $\frac{\lambda}{J}$ close to the isotropic limit ($XXX$ interaction) and in the easy-plane anisotropic phase ($|\Delta|<1$), and superexponential scaling in the easy-axis anisotropic phase ($|\Delta|>1$) within and beyond small perturbations in $\frac{\lambda}{J}$. We also considered the best relative error which can decrease with the particle number faster that shot-noise only for the estimation of $\Delta$ in the isotropic limit and for easy-plane interactions with incommensurate $\frac{\eta}{\pi}$. This scaling identifies regimes for high precision estimation of $\Delta$, which is useful as a measure of bulk properties. Moreover, $\Delta$ can be interpreted as the phases of generalised non-linear interferometers \cite{Luis2004,Boixo2007,Roy2008,Napolitano2011}, where the phase to be estimated is proportional to the strength of the inter-particle interaction in the chain.

The established enhanced metrological performances are not due to entanglement but are a different effect of many-body interactions out of equilibrium. Indeed, in the perturbative regime $\frac{\lambda}{J}\ll1$, relevant for metrological purposes, the NESS is a perturbation of the completely mixed state, thus unentangled since the completely mixed state is in the interior of the set of separable states \cite{BengtssonZyczkowski2006}.

Generalizations to more general hamiltonians, like $XYZ$ interactions, or more general dissipation would be desirable. In the present framework, this requires the derivation of the corresponding NESS. Also, we stress that our results pertain to a completely integrable system with an exactly solvable NESS. Quantum metrology using NESS of generic, non-integrable systems is an open problem left for future research.

\appendix

\section{Jordan decomposition of the transfer matrix} \label{app}

In this appendix we derive the Jordan decomposition of the transfer matrix restricted to the basis $\{|L\rangle,|R\rangle,|1\rangle,|2\rangle,\dots,|d\rangle\}$. First, let's rewrite the restricted transfer matrix as

\begin{equation} \label{Td}
T^{(d)}=|L\rangle\langle L|+|R\rangle\langle R|+\frac{|L\rangle\langle 1|+|1\rangle\langle R|}{2}+{T^{(d)}}',
\end{equation}
where ${T^{(d)}}'$ is the remainder of \eqref{transfer}. The Jordan canonical form of $T^{(d)}$ is

\begin{equation} \label{jordan}
T_J^{(d)}=(V^{(d)})^{-1} T^{(d)} V^{(d)}=
\begin{pmatrix}
\begin{matrix}
1 & 1 \\
0 & 1 
\end{matrix}
& \\
&
\begin{matrix}
\tau_1 & & \\
& \ddots &  \\
& & \tau_d
\end{matrix}
\end{pmatrix},
\end{equation}
where not shown entries are zeros. $\tau_j$ are both eigenvalues of ${T^{(d)}}'$ with eigenvectors $|\tau'_j\rangle$ and eigenvalues of $T^{(d)}$ with eigenvectors $|\tau_j\rangle=\frac{\langle1|\tau'_j\rangle}{2\tau_j-2}|L\rangle+|\tau'_j\rangle$, arranged in decreasing order $|\tau_1|\geqslant\dots\geqslant|\tau_d|$. $T^{(d)}$ also has the eigenvalue $1$ with eigenvector $|L\rangle$ and a defective eigenvector

\begin{eqnarray}
|\psi\rangle & = & \psi_R|R\rangle+\sum_{k=1}^d\psi_k|k\rangle, \nonumber \\
\psi_R & = & \frac{2(d+1)}{d}\left(1-T^{(d)}_{1,1}\right)=\frac{2(d+1)}{d}(1-\Delta^2), \nonumber \\
\psi_k & = & \frac{2(d-k+1)}{d-k+2}\cdot\frac{T^{(d)}_{k,k-1}}{1-T^{(d)}_{k,k}} \, \psi_{k-1}, \quad \psi_1=2 \label{defective}
\end{eqnarray}
(see appendix \ref{app2}). The matrix $V^{(d)}$ is defined by arranging the eigenvectors of $T^{(d)}$ in its columns:

\begin{equation} \label{Vd}
V^{(d)}=
\begin{pmatrix}
1 & 0 & \frac{\langle1|\tau'_1\rangle}{2\tau_1-2} & \dots & \frac{\langle1|\tau'_d\rangle}{2\tau_d-2} \\
0 & \psi_R & 0 & \dots & 0 \\
\begin{matrix}
0 \\
\vdots \\
0
\end{matrix} &
\begin{matrix}
\psi_1 \\
\vdots \\
\psi_d
\end{matrix} &
\begin{bmatrix}
\\
\\
|\tau'_1\rangle \\
\\
\\
\end{bmatrix} & \dots &
\begin{bmatrix}
\\
\\
|\tau'_d\rangle \\
\\
\\
\end{bmatrix}
\end{pmatrix},
\end{equation}
where we write the eigenvectors $|\tau'_j\rangle$ of ${T^{(d)}}'$ inside the squared brackets.

If $\eta=\frac{q\pi}{p}$ with coprime integers $p,q$ the transitions $\big\langle|p|\pm1\big|T\big||p|\big\rangle$ vanish, and we can use the restriction $T^{(d)}$ with $d=|p|-1$ for the computation of the Fisher information. There is no loss of generality to consider coprime $p,q$, since any rational $\frac{\eta}{\pi}$ can be reduced to lowest terms. On the other hand, if $\frac{\eta}{\pi}$ is irrational none of the transitions $\langle j\pm1|T|j\rangle$ vanish, and thus the full matrix $T^{(d)}$ with $d=\lfloor\frac{n}{2}\rfloor$ must be considered in the computations. In both cases, we can analyse the eigenvalues $\tau_j$ by means of the following matrix equality:

\begin{eqnarray}
\!\!\!\!\!\!\!\!\!\! && P^{-1}\left(\mathbbm{1}-{T^{(d)}}'\right)P=\textnormal{sign}(1-\Delta^2)DAD, \qquad \textnormal{with} \nonumber \\
\!\!\!\!\!\!\!\!\!\! && P=|L\rangle\langle L|+|R\rangle\langle R|+\sum_{k=1}^d\left|\frac{\sin(\eta)}{\sin(k\eta)}\right||k\rangle\langle k|, \nonumber \\
\!\!\!\!\!\!\!\!\!\! && D=\sum_{k=1}^d|\sin(k\eta)|\cdot|k\rangle\langle k|, \nonumber \\
\!\!\!\!\!\!\!\!\!\! && A=\sum_{k=1}^d|k\rangle\langle k|-\frac{1}{2}\sum_{k=1}^{d-1}\Big(|k\rangle\langle k+1|+|k+1\rangle\langle k|\Big).
\end{eqnarray}
$A$ is a Toeplitz tridiagonal matrix, thus with eigenvalues $1-\cos\left(\frac{j\pi}{d+1}\right)$ for $j=1,\dots,d$ \cite{Kulkarni1999}. The positivity condition $A>0$ implies

\begin{equation}
\begin{cases}
\displaystyle \mathbbm{1}-{T^{(d)}}'>0\Longleftrightarrow|\tau_j|<1 & \textnormal{if } |\Delta|<1 \\
\displaystyle \mathbbm{1}-{T^{(d)}}'<0\Longleftrightarrow|\tau_j|>1 & \textnormal{if } |\Delta|>1
\end{cases}.
\end{equation}
Note that for $\eta=\frac{q\pi}{p}$ and $d\geqslant|p|$ the matrix $P$ is singular, and indeed the largest of the eigenvalues $\tau_j$ equals $1$.

The Jordan canonical form of powers of the restricted matrix $T^{(d)}$ is used in the computation of the Fisher information:

\begin{equation} \label{powerT}
(T^{(d)}_J)^k=
\begin{pmatrix}
\begin{matrix}
1 & k \\
0 & 1 
\end{matrix}
& \\
&
\begin{matrix}
\tau_1^k & & \\
& \ddots &  \\
& & \tau_d^k
\end{matrix}
\end{pmatrix}.
\end{equation}

\section{Computation of the continuous fractions $C_k$} \label{app2}

The defective eigenvalue of the transfer matrix $T^{(d)}$ \eqref{transfer} is the solution of the the equation

\begin{equation}
(T^{(d)}-\mathbbm{1})|\psi\rangle=|L\rangle.
\end{equation}
Using the knowledge of the entries of $T^{(d)}$, the latter equation can be recast in the following recurrence relation \cite{Prosen2013b}

\begin{eqnarray}
|\psi\rangle & = & \psi_R|R\rangle+\sum_{k=1}^d\psi_k|k\rangle, \nonumber \\
\psi_R & =& 4C_{d-1}\left(1-T^{(d)}_{1,1}\right), \nonumber \\
\psi_k & = & \frac{T^{(d)}_{k,k-1}\psi_{k-1}}{C_{d-k}\left(1-T^{(d)}_{k,k}\right)}, \qquad \psi_1=2,
\end{eqnarray}
with $C_0=1$ and $C_k=1-\frac{1}{4C_{k-1}}$. Note that the transfer matrix considered here is different but similar to the one in \cite{Prosen2013b}, which is

\begin{eqnarray}
&& P^{-1}TP, \qquad \textnormal{with} \nonumber \\
&& P=|L\rangle\langle L|+|R\rangle\langle R|+\sum_{k=1}^d\left|\frac{\sin(\eta)}{\sin(k\eta)}\right||k\rangle\langle k|.
\end{eqnarray}
However, the components of the defective eigenvalues has the same recurrence structure \footnote{We corrected the wrong formula of $\psi_R$ written in \cite{Prosen2013b}. However, the results of \cite{Prosen2013b} still hold true because whenever $\psi_R$ was evaluated the right formula was actually used.}.

The coefficients $C_k$ are continued fractions of the form

\begin{equation} \label{cont.frac}
C_k=a_0+\frac{1}{\displaystyle a_1+\frac{1}{\displaystyle a_2+\frac{1}{\displaystyle \ddots+\frac{1}{a_k}}}},
\end{equation}
with $a_{2j}=1$ and $a_{2j+1}=-4$ $\forall \, j\in\mathbbm{N}$. See \cite{JonesThron} for a reference on continued fractions. Equation \eqref{cont.frac} is equal to

\begin{equation} \label{ratio.cont.frac}
C_k=\frac{f_k}{g_k}
\end{equation}
with

\begin{eqnarray}
f_k=a_k f_{k-1}+f_{k-2}, & \qquad f_{-1}=1, & \quad f_{-2}=0, \nonumber \\
g_k=a_k g_{k-1}+g_{k-2}, & \qquad g_{-1}=0, & \quad g_{-2}=1. \label{recurs}
\end{eqnarray}
The first recursive equation in \eqref{recurs} can be written as

\begin{eqnarray}
\!\!\!
\begin{pmatrix}
f_{2k} \\
f_{2k+1}
\end{pmatrix} & = &
\begin{pmatrix}
1 & 1 \\
-4 & -3
\end{pmatrix}
\begin{pmatrix}
f_{2k-2} \\
f_{2k-1}
\end{pmatrix}=\begin{pmatrix}
1 & 1 \\
-4 & -3
\end{pmatrix}^{n+1}
\begin{pmatrix}
f_{-2} \\
f_{-1}
\end{pmatrix} \nonumber \\
& = & (-)^n\begin{pmatrix}
2k+1 & k+1 \\
-4k-4 & -2k-3
\end{pmatrix}
\begin{pmatrix}
0 \\
1
\end{pmatrix} \nonumber \\
& = &
\begin{pmatrix}
k+1 \\
-2k-3
\end{pmatrix}.
\end{eqnarray}
Similarly for the second recursive relation in \eqref{recurs}

\begin{eqnarray}
\!\!\!
\begin{pmatrix}
g_{2k} \\
g_{2k+1}
\end{pmatrix} & = &
\begin{pmatrix}
1 & 1 \\
-4 & -3
\end{pmatrix}
\begin{pmatrix}
g_{2k-2} \\
g_{2k-1}
\end{pmatrix}=\begin{pmatrix}
1 & 1 \\
-4 & -3
\end{pmatrix}^{n+1}
\begin{pmatrix}
g_{-2} \\
g_{-1}
\end{pmatrix} \nonumber \\
& = & (-)^n\begin{pmatrix}
2k+1 & k+1 \\
-4k-4 & -2k-3
\end{pmatrix}
\begin{pmatrix}
1 \\
0
\end{pmatrix} \nonumber \\
& = &
\begin{pmatrix}
2k+1 \\
-4k-4
\end{pmatrix}.
\end{eqnarray}

From the ratio \eqref{ratio.cont.frac}, we compute $C_k=\frac{k+2}{2k+2}$. Thus, the components of $|\psi\rangle$ become

\begin{eqnarray}
\psi_R & = & \frac{2(d+1)}{d}\left(1-T^{(d)}_{1,1}\right), \nonumber \\
\psi_k & = & \frac{2(d-k+1)}{d-k+2}\cdot\frac{T^{(d)}_{k,k-1}}{1-T^{(d)}_{k,k}} \, \psi_{k-1},
\end{eqnarray}
as in equation \eqref{defective}.

{\bf Acknowledgments.} The work has been supported by grants P1-0044 and J1-5439 of Slovenian Research Agency.

\bibliographystyle{apsrev4-1}

\bibliography{Fisher_NESS-ref}

\end{document}